\documentstyle[a4,12pt,epsf]{article}
\newcommand{\noi}{\noindent}
\newcommand{\eq}{\begin{equation}}
\newcommand{\en}{\end{equation}}
\newcommand{\eqa}{\begin{eqnarray}}
\newcommand{\ena}{\end{eqnarray}}

\newcommand{\vp}{{\vec p}}

\newcommand{\vq}{{\vec q}}
\newcommand{\vqp}{{\vec q}_{\perp}}
\newcommand{\vx}{{\vec x}}
\newcommand{\vxp}{{\vec x}_{\perp}}

\newcommand{\vR}{{\vec R}}

\newcommand{\tOmega}{{\tilde \Omega}}

\newcommand{\cak}{ {\cal K}}

\newcommand{\bt}{{\bar \theta}}

\newcommand{\bpartial}{{\bar \partial}}

\newcommand{\lra}{\longrightarrow}

\begin{document}
\hbox{}
\noindent July 1996 \hfill HU Berlin--EP--96/31

\vspace{0.5cm}
\begin{center}

\renewcommand{\thefootnote}{\fnsymbol{footnote}}
\setcounter{footnote}{0}

{\LARGE Classical solutions in lattice gauge theories}
\footnote{Work supported by the Deutsche
Forschungsgemeinschaft under research grant Mu 932/1-4} \\

\vspace*{0.5cm}
{\large
V.K.~Mitrjushkin $\mbox{}^1$\footnote{Permanent adress:
Joint Institute for Nuclear Research, Dubna, Russia}
}\\

\vspace*{0.2cm}
{\normalsize
$\mbox{}^1$ {\em Institut f\"{u}r Physik, Humboldt-Universit\"{a}t,
10115 Berlin, Germany}\\     
}     
\vspace*{0.5cm}
{\bf Abstract}
\end{center}

The solutions of the classical equations of
motion on a periodic lattice are found which correspond to
abelian single and double Dirac sheets.
These solutions exist also in non--abelian theories.
Possible applications of these solutions to the calculation of 
gauge dependent and gauge invariant observables 
are discussed.

\renewcommand{\thefootnote}{\arabic{footnote}}
\setcounter{footnote}{0}

\section{Introduction}

The lattice approach gives the possibility to use powerful numerical
methods to calculate gauge invariant as well as gauge dependent objects 
\cite{wil}. 
However very often the interpretation of the results of numerical
calculations needs analytical methods on a lattice, e.g. perturbation theory.
The importance of certain classical configurations for understanding
of the nonperturbative physics was stressed in \cite{pol1}.
The very existence of non--zero solutions of the classical
equations of motion can explain some effectes observed
in simulations and provide insight into the
relation between continuum and lattice theories.
The perturbative expansion about the non--zero solutions of the
classical equations of motion permits to take into account 
nonperturbative contributions, at least, partially.

This work is devoted to the study of the classical solutions on 
a periodic lattice and some applications of these solutions.
The special accent in this work is made on the discussion
of gauge dependent objects, e.g. photon correlator, especially
in the connection with the Gribov ambiguity problem \cite{grib}.
 
In many practical situations in lattice gauge theory it is rather useful
to calculate gauge variant quantities, e.g., fermionic and gauge field
correlators.  For example, calculating the gauge variant gluon
correlators one can attempt to obtain information about gauge invariant
observables, like energies and masses.
The study of the Gribov problem  on a lattice
has demonstrated the existence of the gauge--fixing ambiguities
both in abelian and nonabelian theories.
The non-uniqueness of the solutions of the commonly used gauge
conditions, e.g.  Lorentz gauge, can strongly affect the numerical
computation of gauge dependent quantities.

Let $~S(U)~$ be the lattice action with gauge symmetry group $~G$.
Under the gauge transformations $V_x$ the link variables
$~U_{x \mu}~$ transform 
$~ U_{x \mu} \stackrel{V}{\lra} U_{x \mu}^{V} \equiv
V_x  U_{x \mu} V_{x+\mu}^{\dagger}$,
$~U_{x\mu},V_x \in G.$
The partition function $Z$ is

\eq
Z \equiv \int \! \prod_{x\mu} \! dU_{x \mu} \, e^{- S(U)}~,
                                          \label{part_f_1}
\en

\noi where $dU_{x\mu}$ is the Haar measure.  The usual way to fix the
gauge is to insert in the functional integral in eq.(\ref{part_f_1})
the identity
$~1 = J(U) \int \! \prod_{x} \! dV_x  \, 
\prod_x \delta \Bigl( F_x(U^{V}) \Bigr)$, where
$~F_x(U)$ is a gauge fixing functional.
The Faddeev--Popov determinant $J(U)$ is 

\eq
J^{-1}(U) = \sum_{V^{\prime}} \det \mbox{}^{-1} 
\left( \frac{\partial F_{x}^{V }}{\partial V_y}
 \Big|_{V =V^{\prime}} \right) ~,
\en

\noi where $V^{\prime}$ satisfies $~F_{x}(U^{V^{\prime}}) =0$.
The average of any (gauge invariant or not) functional $~\Phi~$
is defined as

\eq
\langle \Phi (U) \rangle \equiv Z^{-1} \int \! \prod_{x\mu} \! dU_{x\mu}~
J(U) \cdot \prod_x \delta \Bigl( F_x(U) \Bigr) \cdot
\Phi (U) \cdot e^{-S(U)}~.
                \label{func_1}
\en

\noi Assuming that there are no gauge copies
one can represent eq.(\ref{func_1}) in the following form

\eq
\langle \Phi (U) \rangle = Z^{-1} \int \! \prod_{x\mu} 
\! dU_{x\mu}~ \Phi (U^{V^0}) \cdot e^{-S(U)} ~,
               \label{phi_gf}
\en

\noi where $V^0 \equiv \bigl\{ V^0_x \bigr\}$ is the gauge
transformation for which $~F(U^{V^0}) = 0$.
This means that the expected value of any functional $~\Phi (U)~$
may be calculated from the ensemble of configurations weighted
without gauge constraint, but that each configuration should be
gauge transformed into $~F(U) =0~$ gauge before evaluating
$~\Phi$. 
Only if the field $U$ has no gauge copies, i.e.,
$~ F_{x}(U^{V^{\prime}})~$ equals to zero for only one 
value of the gauge parameter $V^{\prime}$,
then $J(U)$ is determined by fluctuations about a single
field $~U^{V^{\prime}}$, and eq.(\ref{phi_gf}) provides
an unambigous definition of the average functional 
$\langle \Phi \rangle$.

On the perturbative language the existence of the gauge copies can be
interpreted as a problem of the gauge copies of the `vacuum'
configurations, i.e.  the solutions of the classical equations of
motion.  Assuming that every configuration is some small fluctuation
about the corresponding `vacuum' one can find all gauge copies of this
configuration provided all gauge copies of the `vacuum' are known.

Throughout this work the $~4d~$ lattice 
with periodic boundary conditons is considered.
$N_{\mu}$ is the lattice size in the direction $~\mu$, and
$~V_4 =N_1 N_2 N_3 N_4$.
The lattice derivatives are
$~\partial_{\mu} f(x) = f(x+{\hat \mu}) -f(x)~$ and
$~\bpartial_{\mu} f(x) = f(x) - f(x-{\hat \mu})$,
and the lattice spacing is chosen to be unity.

\section{Classical solutions in the compact $U(1)$ theory}

In the case of the $U(1)$ gauge group 
$U_{x\mu} = e^{i\theta_{x\mu}} \in U(1)$,
and the Wilson action $~S_W(U)~$ is

\eq
S_W(U) = \frac{1}{g^2} \sum_{x} \sum_{\mu \nu =1}^4
        \,  \bigl( 1 - \cos \theta_{x;\mu \nu} \bigr) ~;
\quad \beta = \frac{2}{g^2}~,
\en

\noi where $~\theta_{x;\mu\nu} 
= \partial_{\mu} \theta_{x\nu} - \partial_{\nu} \theta_{x\mu}~$
are the plaquette angles.
It is important to note that all link angles $\theta_{x\mu}$ are
compact variables  ($ -\pi < \theta_{x\mu} \le \pi $), and the
gauge transformations 
$\theta_{x\mu} \stackrel
{\Omega}{\to} \theta_{x\mu} - \partial_{\mu} \Omega_x$
 are understood modulo $2\pi$.

The plaquette angle $~\theta_P \equiv \theta_{x;\, \mu \nu}~$
can be split up: $~\theta_P = [\theta_P] + 2\pi n_P$, where
$~[\theta_{P}] \in (-\pi ;\pi ]$ and $n_P = 0, \pm 1, \pm 2$.
The plaquettes with $~n_P \neq 0~$ are called Dirac plaquettes.
The dual integer valued plaquettes $~m_{x,\mu \nu} =
\frac {1}{2} \varepsilon_{\mu \nu \rho \sigma} n_{x,\rho \sigma}~$
form Dirac sheets \cite{dgt}.

The classical equations of motion are

\eq
\frac{\partial S_W (U)}{\partial \theta_{x \mu}} 
\Bigl|_{\theta =\theta^{cl.}}
=\frac{2}{g^2} \sum_{\nu } \bpartial_{\nu}
\sin \theta^{cl.}_{x;\mu \nu} = 0~.
                  \label{class_eq_u1}
\en

\noi The $x$--independent solutions of these equations
$\theta^{cl.}_{x\mu}=\phi_{\mu}$ (zero--momentum modes) and their influence on
the gauge--dependent correlators were discussed in \cite{vkm}.

Evidently, any function $\bt_{x\mu}$ is the solution of
the eq.(\ref{class_eq_u1}) if the corresponding plaquette
angle $~\bt_P~$ satisfies the following
condition

\eqa
\bt_{P} =
   \left\{ \begin{array}{cc}
  2\pi + \Delta & \quad \mbox{if} \qquad P = P^{\ast}~;         \\
         \Delta & \quad \mbox{if} \qquad P \ne P^{\ast}~,
   \end{array}  \right.
               \label{dirac_cond_1}
\ena

\noi where $~P^{\ast}~$ is some plaquette in the plane 
$~(x_{\rho};x_{\sigma})$, and $\theta_{x\mu} =0$ for $\mu \ne \rho ;\sigma$.
Periodicity presumes 

\eq
\sum_{P\in (x_{\rho};x_{\sigma})} \bt_P =0,
\qquad \Delta = -\frac{2\pi}{N_{\rho}N_{\sigma}}~.
               \label{dirac_cond_2}
\en

\noi Therefore, $P^{\ast}$ is a Dirac plaquette.

To derive the explicit solution of the eq.(\ref{class_eq_u1})
satisfying the condition in eq.(\ref{dirac_cond_1})
it is convenient to start with a $4d$ volume in the continuum. 
Let us choose $(\rho ;\sigma )=(1;2)$, and 
let $B^{cont}_3 (\vxp )$ be the third component of the magnetic field :
$$B^{cont}_3 (\vxp )=H 
\Bigl[ \delta (\vxp - \vxp^{(1)}) -\frac{1}{N_1N_2} \Bigr],
\qquad \vxp =(x_1;x_2), $$
\noi where $\vxp^{(1)}$ denotes the position of the string.
The total flux through the plane $(x_1;x_2)$ equals to zero.
It is an easy exercise 
to find the corresponding vector--potential $A^{cont}_{\mu}(x)$.
Embedding the continuum solution into the lattice one can define
the link variables $\theta_{x\mu}$ via an integral of the $gA^{cont}_{\mu}(x)$
along the corresponding links. The constant $H$ is defined to fulfill
eq's.(\ref{dirac_cond_1},\ref{dirac_cond_2}).
Choosing the Lorentz (or Landau) gauge

\eq
\sum_{\mu} \bpartial_{\mu} \theta_{x\mu} =0~,
               \label{gauge_cond_1}
\en

\noi one arrives after some algebra at the expression for the 
single Dirac sheet

\eqa
\bt_{x1}(\vR ) &=& \frac{2\pi i }{N_1N_2}
\sum_{\vqp \ne 0} \frac{\cak_2} {{\vec \cak }^{\, 2}_{\perp}}
\cdot e^{i\vqp (\vxp - \vR ) - \frac{i}{2}q_2 }~;
                 \label{sds_u1}
\\
\nonumber \\
\bt_{x2}(\vR ) &=& -\frac{2\pi i }{N_1N_2}
\sum_{\vqp \ne 0} \frac{\cak_1} {{\vec \cak }^{\, 2}_{\perp}}
\cdot e^{i\vqp (\vxp - \vR ) - \frac{i}{2}q_1 }~;
\nonumber \\
\nonumber \\
q_j &=& \frac{2\pi}{N_j} n_j;
\quad n_j=-\frac{1}{2}N_j+1;~\ldots ~;\frac{1}{2}N_j~,
\nonumber 
\ena

\noi where $~\cak_{\mu} =2\sin \frac{q_{\mu}}{2}$ and
$~{\vec \cak}^2_{\perp} = \cak_1^2+\cak_2^2$. 
The twodimensional vector $\vR =(R_1;R_2)$ corresponds to
the position of the Dirac plaquette in the $(x_1;x_2)$ plane:
$$~\bt_{x;12} = 2\pi \cdot \delta_{\vxp ;\vR} -\frac{2\pi}{N_1N_2}~.$$
Of course, $~\bt^{\prime}_{x\mu} =\phi_{\mu} + \bt_{x\mu}~$ is also
the solution of eq's.(\ref{class_eq_u1},\ref{gauge_cond_1}).

It is easy to check that the single Dirac sheet solution 
$\bt_{x\mu}(\vR )$ corresponds
to the local minimum of the action, i.e. it is stable with respect 
to small fluctuations. The existence of the long--living
metastable states corresponding to single Dirac sheets 
was observed in simulations in the pure gauge $U(1)$ theory
\cite{neu}.

In Figure \ref{fig:kink}a(b) the dependence of the first
component $\bt_{x;1}(\vR )$ on $x_2$($x_1$) is shown for 
different values $x_1(x_2)$. 
It demonstrates a characteristic kink--like
behavior along the axis $x_2$.
The second component $\bt_{x;2}$ shows the similar behavior 
along the axis $x_1$.

The classical gauge action $S_{cl.}=S(\bt )$ is

\eqa
S(\bt ) =\frac{2V_4}{g^2} \Bigl( 1-\cos \Delta \Bigr)~.
\ena

\noi The value $S(\bt )$ depends on the geometry of the lattice.
Let us choose $N_s=N_1=N_2=N_3 \to \infty$.
On a symmetric lattice $N_4=N_s$ the action $S(\bt )$ is
non--zero and finite :
$$S(\bt )=\frac{4\pi^2}{g^2} < \infty~.$$
\noi In the zero--temperature limit, i.e. $N_4 \to \infty$, $N_s/N_4 \to 0$,
the classical action is
$$S(\bt )=\frac{4\pi^2}{g^2} \frac{N_s}{N_4} \to 0~,$$ 
\noi if the Dirac plaquette is time--like, i.e. 
$(\rho ;\sigma)=(4;i),~i=1,2,3$.
In the finite--temperature limit, i.e. $N_4/N_s \to 0$, the action is
$$S(\bt )=\frac{4\pi^2}{g^2} \frac{N_4}{N_s} \to 0~,$$ 
\noi if the Dirac plaquette is space--like.

Another possible solution of the classical equation of motion
-- double Dirac sheet  -- consists of the two single Dirac
sheets with an opposite orientation of the flux :

\eq
\bt_{xi}(\vR^{a};\vR^{b} ) = \bt_{xi}(\vR^{a}) - \bt_{xi}(\vR^{b}),
\qquad i=1;2~,
                \label{dds}
\en 
 
\noi where vectors $\vR^{a}$ and $\vR^{b}$ correspond to the two 
Dirac plaquettes in the plane $(x_1;x_2)$. It is easy to see that

\eq
\bt_{x;12}(\vR^a;\vR^b ) =2\pi \cdot \Bigl[ \delta_{\vxp ;\vR^a}
-\delta_{\vxp ;\vR^b} \Bigr]~.
\en 

\noi The double Dirac sheet $~\bt_{x;i}(\vR^a;\vR^b )~$ has a
zero action: $~S(\bt ) = 0$. 

Gauge transformations can shift the Dirac sheets and change their form.
For example, the `big' gauge transformation function $~\Omega_x~$

\eqa
\Omega_{x} &=& -\frac{2\pi }{N_1N_2} \sum_{\vqp \ne 0} 
\frac{ e^{i\vqp (\vxp - \vR) }}{{\vec \cak }^{\, 2}_{\perp} } 
\cdot \Bigl( 1 - e^{-iq_2 } \Bigr)~;
                 \label{shift_x}
\\
\nonumber \\
\Box \Omega_x &=& 2\pi \cdot 
\Bigl[ \delta_{\vxp ;\vR} -\delta_{\vxp ;\vR -{\hat 1}} \Bigr]~,
\qquad
\Box = \sum_{\mu} \bpartial_{\mu}\partial_{\mu}~,
\nonumber
\ena

\noi shifts the single Dirac sheet in the $x_1$--direction :

\eqa
\bt_{x1}^{\Omega}(\vR ) &=& \bt_{x1}(\vR - {\hat 1} )~;
\nonumber \\
\nonumber \\
\bt_{x2}^{\Omega}(\vR) &=& \bt_{x2}(\vR -{\hat 1} ) 
- \Bigl[ 2\pi \cdot \delta_{ \vxp ; \vR } - \frac{ 2\pi } {N_1N_2}
\Bigr].
\ena

\noi The gauge transformation $\Omega_x$ 
in eq.(\ref{shift_x}) applied to
the zero--field creates a double Dirac sheet as  in
eq.(\ref{dds}) with $\vR^a =\vR $ and $\vR^b =\vR -{\hat 1}$ :

\eqa
\partial_1 \Omega_{x} &=& -\bt_{x1}(\vR;\vR -{\hat 1})~; 
\nonumber \\
\nonumber \\
\partial_2 \Omega_{x} &=& -\bt_{x2}(\vR;\vR -{\hat 1}) -
2\pi \cdot \delta_{\vxp ;\vR }+\frac{2\pi }{N_1N_2}~.
\nonumber 
\ena

\noi Therefore,  $~\bt_{x;i}(\vR^a;\vR^b )$ 
is a Gribov copy of the zero solution $~\theta^{cl.}_{x\mu} =0$.

It is not difficult now to obtain the general Dirac sheet 
solutions, i.e. the Dirac sheets curved in the fourdimensional
space.
As an example, let us define the `big' gauge transformation 
$~\tOmega_x~$ depending on $~(\vxp ;x_3)$ :

\eq
\tOmega_{x}  = \delta_{x_3;\xi_3} \cdot \Omega_x~,
\en

\noi where $\Omega_x$ is defined in eq.(\ref{shift_x}), and
$\xi_3$ is some number.  After gauge transformation

\eq
\bt_{x\mu}(\vR ) \stackrel{\tOmega}{\lra} \bt^{\tOmega}_{x\mu} 
= \bt_{x\mu}(\vR ) - \partial_{\mu} \tOmega_{x}~,
               \label{omega_tilde_1}
\en

\noi the flat Dirac sheet $~\bt_{x\mu}(\vR ) ~$ (Figure
\ref{fig:nonflat}a) changes its shape as shown
in Figure \ref{fig:nonflat}b.
The new field $\bt^{\Omega}_{x\mu}$ does not fulfill the Lorentz gauge :
$~\sum_{\mu} \bpartial_{\mu} \bt^{\Omega}_{x\mu} 
= - \Omega_x \cdot \bpartial_{3}\partial_{3} \delta_{x_3;\xi_3} \ne 0$.
To restore the Lorentz gauge one should make an additional
gauge transformation 
$~\bt^{\tOmega}_{x\mu}(\vR ) \stackrel{\omega}{\lra} 
\bt^{\tOmega}_{x\mu}(\vR ) -\partial_{\mu} \omega_{x}$, where

\eq
\omega_x = \frac{2\pi i}{V_3}\sum_{\vqp \ne 0} \sum_{q_3}
e^{i\vqp (\vxp -\vR )+iq_3(x_3-\xi_3)-\frac{i}{2}q_2} 
\cdot \frac{\cak_2 \cak_3^2} {{\vec \cak}^2{\vec \cak}^2_{\perp}}~,
\en

\noi where ${\vec \cak}^{\, 2} = {\vec \cak}^{\, 2}_{\perp} + \cak^2_3$.
Therefore, the successive application of the `big' gauge transformations
creates a Dirac sheet of any possible geometry in the $4d$ space.

The symbolic representation of the dependence of the action
$S(U)$ on the gauge field configuration $\{ U_{x\mu} \}$ 
is shown in Figure \ref{fig:vacuum}.
The absolute minima correspond to double
Dirac sheets, and the local minima correspond to single
Dirac sheets. Of course, there are also the local minima
corresponding to the two single Dirac sheets, etc. (are not shown).
In principle, the existence of non--trivial classical solutions of other type
can not be excluded.

It is interesting to discuss the role of the lattice
(DeGrand--Toissaint) monopoles \cite{dgt} in connection with the
stability of the Dirac sheet solutions.  The breaking of a
periodically closed single Dirac sheet means the appearence of 
monopole--antimonopole pairs. Therefore, the tunneling from one `vacuum'
to another, i.e. the changing of the number of the single Dirac sheets,
is accompanied by the creation of monopoles. 
The modification of the compact action which suppresses the monopoles
\cite{mon1,mon2} prevents the breaking of the sheet and makes the 
tunneling impossible.
In this case the choice of the initial configuration in numerical
calculations will correspond to the choice of the `vacuum'.

It is easy to see that the single and double Dirac sheets
are also the classical solutions of the non--abelian theories. 
For example, in the case of the $SU(2)$ gauge group the link variables

\eq
{\bar U}_{x;\mu}(\vR_1;\vR_2) \equiv
\exp \Bigl\{ i\bt_{x\mu}(\vR_1;\vR_2 ) \cdot \sigma_3 \Bigr\} \in SU(2) 
\en

\noi satisfy evidently the classical equations of motion and
correspond to the zero action.

\section{The gauge--dependent photon correlator}

As an example of the importance of the classical solutions 
$\theta^{cl.}_{x\mu} \ne 0$ in lattice calculations let us consider
the gauge--dependent photon correlator $~\Gamma_{\mu}(\tau ;\vp )~$ 

\eqa
\Gamma_{\mu} (\tau ;\vp ) &=& \Bigl\langle 
{\cal O}_{\mu}^{\ast}(\tau ;\vp ) {\cal O}_{\mu}(0;\vp ) \Bigr\rangle
= \frac{1}{N_4}\sum_{t=0}^{N_4-1} \Bigl\langle 
{\cal O}_{\mu}^{\ast}(t \oplus \tau ;\vp ) {\cal O}_{\mu}(t;\vp ) \Bigr\rangle~;
\nonumber \\
\nonumber \\
{\cal O}_{\mu}(\tau ;\vp ) &=& \sum_{\vx} e^{-i\vp \vx -\frac{i}{2}p_{\mu}} 
\cdot \sin \theta_{x\mu}~,
\qquad \mu =1,2,3~,
\ena

\noi where $t \oplus \tau = t+\tau \mbox{ mod } N_4$.
Evidently, $~\langle O_{\mu} \rangle =0$.

In \cite{nak} it was shown that in the Coulomb phase some of the
gauge copies produce a photon correlator with a decay behavior
inconsistent with the zero mass behavior.
Numerical study \cite{bmmp} has shown that there is a connection
between `bad' gauge copies and the appearence of configurations with
double Dirac sheets. Now we can explain this effect.

Let us choose  the momentum $~\vp =(0;p_2;0)~$ with $~p_2 \ne 0$, and 
$\mu =1$.
The perturbative expansion about the zero solution of the classical
equation of motion $\theta^{cl}_{x\mu} =0$, i.e.  the standard
perturbation theory, gives in the lowest approximation

\eq
\Gamma^{stan.}_1(\tau;\vp ) \sim e^{-\tau E_p }+e^{-(N_4 - \tau )E_p}~,
               \label{cor_1}
\en

\noi where the energy $E_p$ satisfies the lattice dispersion
relation

\eq
\sinh^2 \frac{E_p}{2} = \sum_{i=1}^3 \sin^2 \frac{p_i}{2}~.
               \label{ldr}
\en

The non--zero solutions of the classical equations of motion
$\theta^{cl.}_{x\mu} \ne 0$ have to be taken into account.
Representing the gauge field in the form 
$\theta_{x\mu} =\theta^{cl.}_{x\mu}+gA_{x\mu}$ and expanding in powers
of $g$ one obtains

\eqa
{\cal O}_1(\tau ;\vp ) &=& \Phi (\tau ; \vp ) + \frac{g}{V_3} \sum_{\vq} 
A_1(\tau ;\vq ) \cdot e^{\frac{i}{2}aq_1} \Psi (\tau ;\vp -\vq )
+ O(g^2) ~;
               \label{oper}
\\
\nonumber \\
\Phi (\tau ; \vq ) &=& \frac{1}{V_3}  
\sum_{\vx} e^{-i\vq \vx } \cdot \sin \theta^{cl.}_{x1} ~;
\qquad
\Psi (\tau ; \vq ) = \frac{1}{V_3} 
 \sum_{\vx} e^{-i\vq \vx } \cdot \cos \theta^{cl.}_{x1} ~.
\nonumber 
\ena

\noi In what follows the $O(g^2)$ term in the r.h.s.
in eq.(\ref{oper}) is discarded.
The transverse correlator $\Gamma_1(\tau ;\vp )$ is

\eqa
\Gamma_{1} (\tau ;\vp ) &=& \Gamma^{(0)}_{1} (\tau ;\vp ) 
+ g^2 \cdot \Gamma^{(1)}_{1} (\tau ;\vp ) +\ldots~;
\nonumber \\
\nonumber \\
\Gamma^{(0)}_{1} (\tau ;\vp ) &=& \frac{1}{N_4}\sum_{t=0}^{N_4-1}
\Phi^{\ast}(t \oplus \tau ;\vp ) \Phi(t;\vp ) ~;
\nonumber \\
\nonumber \\
\Gamma^{(1)}_{1} (\tau ;\vp ) &=& \frac{1}{2V_3} \sum_{\vq} G(\tau ;\vq )
\frac{1}{N_4}\sum_{t=0}^{N_4-1} \Psi^{\ast}(\tau \oplus t;\vp -\vq )\Psi(t;\vp -\vq )~,
               \label{cor_2}
\ena

\noi where $G(\tau ;\vp )$ is 

\eqa
G(\tau ;\vq ) &=& \left[ 1 + \frac{\cak_1^2(q)}{2\sinh E_q} 
\cdot \frac{d}{dE_q } \right] G_0(\tau ; \vq )~;
\\
\nonumber \\
G_0(\tau;\vq ) &=& \frac{1}{2 \sinh E_q} 
\frac{1}{1 - e^{-N_4E_q}} \cdot
\Bigl[ e^{-\tau E_q } + e^{-(N_4 - \tau )E_q} \Bigr] +\ldots ~.
\nonumber
\ena

\noi It is easy to see that the expansion about a Dirac sheet solution
gives a contribution to the correlator very different from that in
eq.({\ref{cor_1}).  

As an example, let us choose the flat double Dirac
sheet with the space--like Dirac plaquettes $~\theta^{cl.}_{x\mu} =
\bt_{x\mu}(\vR_1 ;\vR_2 )$ as defined in eq.(\ref{dds}).  In this case
$\Phi$ and $\Psi$ do not depend on $\tau$ :

\eq
\Phi (\vq ) = \frac{\delta_{q_3;0}}{N_1N_2}  
\sum_{\vxp} e^{-i\vqp \vxp } \sin \bt_{x1} ~;
\quad
\Psi (\vq ) = \frac{\delta_{q_3;0}}{N_1N_2}  
\sum_{\vxp} e^{-i\vqp \vxp } \cos \bt_{x1} ~,
\nonumber 
\en

\noi  and the correlator in the double Dirac sheet background is

\eq
\Gamma^{dds}_{1} (\tau ;\vp ) = \Bigl| \Phi (\vp )\Bigr|^2
+ \frac{g^2}{2V_3} \sum_{\vqp} G(\tau ;\vqp + \vp ) 
\cdot \Bigl| \Psi(\vqp ) \Bigr|^2~.
               \label{cor_3}
\en

\noi In Figure \ref{fig:cor_3} the correlator 
$\Gamma^{dds}_{1}(\tau ;\vp )$ is shown in comparison
with the `standard' photon correlator $\Gamma^{stan.}_{1}(\tau ;\vp )$
on a $12^3\times 24$ lattice for $\vp =(0;\frac{\pi}{6};0)$ 
and $\beta =1.1$. Both correlators are normalized to unity at $\tau =0$.
The $\tau$--independent term $~\Bigl| \Phi (\vp )\Bigr|^2~$
in the r.h.s. in eq.(\ref{cor_3}) gives the dominant contribution
which results in a photon correlator inconsistent
with the zero mass behavior. This effect shows rather weak volume
dependence.

\section{Conclusions and discussions}

The solutions of the classical equations of motion on a periodic lattice
are found which correspond to the $U(1)$ single and double Dirac sheets.
These solutions demonstrate the typical kink--like (or kink--antikink)
behavior and have finite energy.
The perturbative expansion should take into account
the contribution of these stationary points.

The double Dirac sheets are the Gribov copies of the trivial solution
$\theta^{cl}_{x\mu} =0$ with zero action.  They can be created from the
zero solution by the `big' gauge--transformations $\Omega_x$.
Therefore, the gauge invariant objects are not affected by the double
Dirac sheets.
On the contrary, the influence of the double Dirac sheets on gauge
dependent values can be of crucial importance, as it was demonstrated
on the example of the photon correlator.

The single Dirac sheets, i.e.  the classical solutions corresponding to
local minima of the action, deserve a special study.  Contrary to
the case of the double Dirac sheets the single Dirac sheets give a
non--zero contribution to gauge invariant observables, e.g.  average
plaquette.  This contribution is of nonperturbative nature, i.e.
$~\sim~\exp \bigl\{ -C/g^2 \bigr\}$, and can mimic the `condensate'
contribution.

The tunneling from one `vacuum' to another is accompanied by the
creation of lattice monopoles.
The proper modification of the action, i.e. the suppression of
monopoles, excludes the tunneling making the `vacua' stable.
In this case the choice of the initial configuration in numerical
calculations will correspond to the choice of the `vacuum'.

The single and double Dirac sheets
are also the classical solutions of the non--abelian theories. 
Therefore, the gauge dependent gluon correlators are supposed to
be strongly influenced by the Dirac sheets as it happens in the
case of the photon correlator.
This question as well as question about the connection of the
Dirac sheets with confinement needs further clarification.
This work is in progress.

\vspace{2cm}
\noi {\large {\bf Figure Captions}}

\vspace{0.25cm}
\noi {\bf Figure 1}~~ The link angle $\theta_{x1}$ as defined in
eq.(\protect{\ref{sds_u1}}) on a lattice with $N_1=N_2=12$.  The lines
are to guide the eyes.

\vspace{0.25cm}
\noi {\bf Figure 2}~~ The flat Dirac string ({\bf a}) and
distorted Dirac string ({\bf b}).

\vspace{0.25cm}
\noi {\bf Figure 3}~~ The symbolic representation of the dependence of
the action $S(U)$ on the gauge field configuration $\{ U_{x\mu} \}$.
The absolute minima correspond to double Dirac sheets, and the local
minima correspond to single Dirac sheets.
There are also the local minima corresponding to the two single
Dirac sheets, etc. (not shown).

\vspace{0.25cm}
\noi {\bf Figure 4}~~ The normalized correlators $\Gamma^{dds}_{1}(\tau
;{\vec p})$ (circles) and $\Gamma_1^{stan.}(\tau ;{\vec p})$ (diamonds)
on a $12^3\times 24$ lattice for $\vp =(0;\frac{\pi}{6};0)$ and
$\beta =1.1$.

%
%
%
\begin{figure}[pt]
\vspace{-1.0cm}
\begin{center}
\leavevmode
\hbox{
\epsfysize=400pt\epsfbox{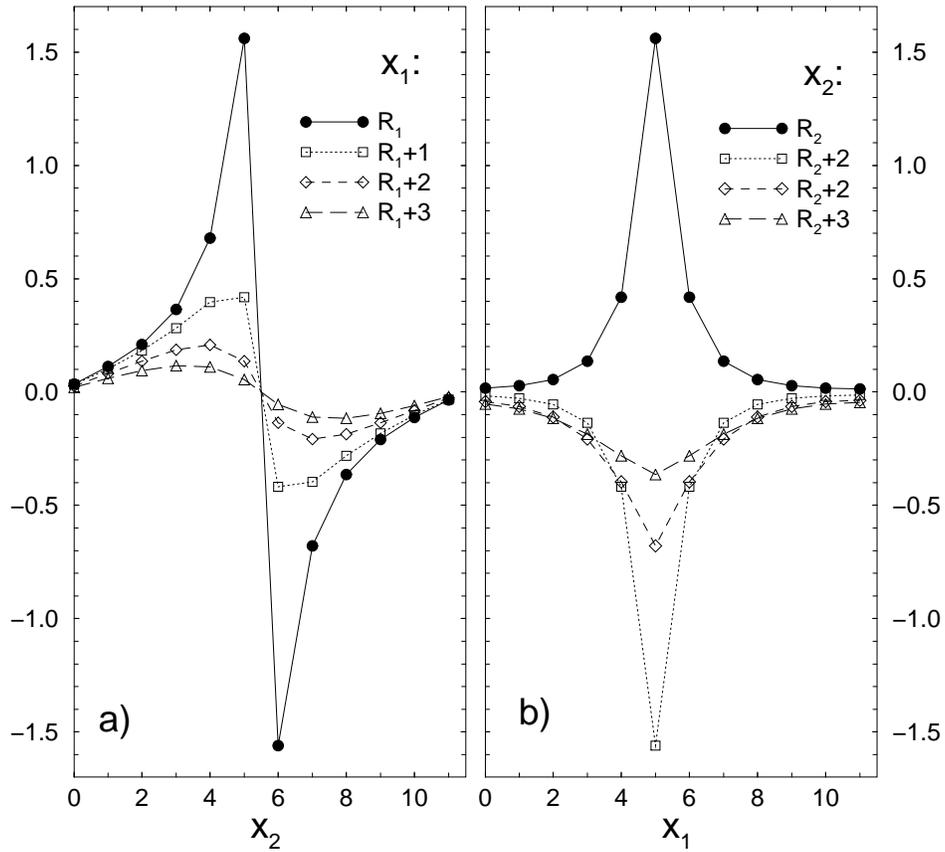}
     }
\end{center}
\vskip 0.25truecm
\caption{
The link angle $\theta_{x1}$ as defined in eq.(\protect{\ref{sds_u1}})
on a lattice with $N_1=N_2=12$.
The lines are to guide the eyes.
}
\label{fig:kink}
\vskip -0.2truecm
\end{figure}

%
%
%
\begin{figure}[pt]
\setlength{\unitlength}{2cm}
\begin{picture}(7,3)
\thinlines
\put(1,2){\line(1,0){1.5}}
\put(1,2){\line(-1,-1){1.0}}
\put(1,2){\line(0,1){3.4}}
\thicklines
\put(0.0,0.8){$x_2$}
\put(2.6,1.9){$x_1$}
\put(1,5.5){$x_3$}
\put(1.0,0.5){{\bf a)}}
\put(1,2){\line(1,0) {1}}
\put(1,2){\line(-1,-1) {0.6}}
\put(2,2){\line(-1,-1) {0.6}}
\put(0.4,1.4){\line(1,0) {1}}
\put(1,3){\line(1,0) {1}}
\put(1,3){\line(-1,-1) {0.6}}
\put(2,3){\line(-1,-1) {0.6}}
\put(0.4,2.4){\line(1,0) {1}}
\put(1,4){\line(1,0) {1}}
\put(1,4){\line(-1,-1) {0.6}}
\put(2,4){\line(-1,-1) {0.6}}
\put(0.4,3.4){\line(1,0) {1}}
\put(1,5){\line(1,0) {1}}
\put(1,5){\line(-1,-1) {0.6}}
\put(2,5){\line(-1,-1) {0.6}}
\put(0.4,4.4){\line(1,0) {1}}
\thinlines
\put(4,2){\line(1,0){1.5}}
\put(4,2){\line(-1,-1){1.0}}
\put(4,2){\line(0,1){3.4}}
\thicklines
\put(3.0,0.8){$x_2$}
\put(5.6,1.9){$x_1$}
\put(4,5.5){$x_3$}
\put(4.0,0.5){{\bf b)}}
\put(4,2){\line(1,0) {1}}
\put(4,2){\line(-1,-1) {0.6}}
\put(5,2){\line(-1,-1) {0.6}}
\put(3.4,1.4){\line(1,0) {1}}
\put(4,3){\line(1,0) {1}}
\put(4,3){\line(-1,-1) {0.6}}
\put(5,3){\line(-1,-1) {0.6}}
\put(3.4,2.4){\line(1,0) {1}}
\put(5,4){\line(1,0) {1}}
\put(5,4){\line(-1,-1) {0.6}}
\put(6,4){\line(-1,-1) {0.6}}
\put(4.4,3.4){\line(1,0) {1}}
\put(5,4){\line(0,1) {1}}
\put(5,3){\line(0,1) {0.41}}
\put(4.4,2.4){\line(0,1) {1}}
\put(4.4,3.4){\line(0,1) {1}}
\put(4,5){\line(1,0) {1}}
\put(4,5){\line(-1,-1) {0.6}}
\put(5,5){\line(-1,-1) {0.6}}
\put(3.4,4.4){\line(1,0) {1}}
\end{picture}\par
\caption{The flat Dirac string ({\bf a}) and distorted
Dirac string ({\bf b}).}
\label{fig:nonflat}
\end{figure}
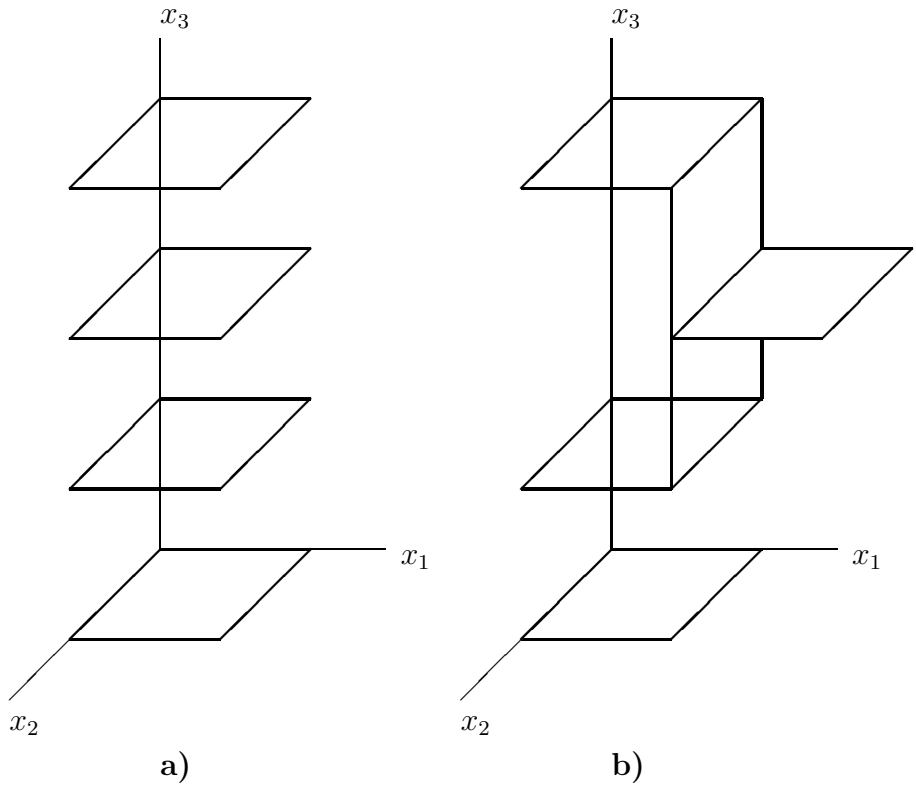

%
\begin{figure}[h]
\setlength{\unitlength}{2cm}
\begin{picture}(7,2)
\thinlines
\put(0,0){\vector(1,0){5.4}}
\put(0,0){\vector(0,1){1.5}}
\thicklines
\put(5.5,-0.25){$\bigl\{ U_{x\mu} \bigr\}$}
\put(0,1.6){S(U)}
\put(0.0,0.5){\oval(1,1)[br]}
\put(0.70,0.5){\oval(0.4,0.5)[t]}
\put(1.10,0.5){\oval(0.4,0.5)[b]}
\put(1.50,0.5){\oval(0.4,0.80)[t]}
\put(2.20,0.5){\oval(1,1)[b]}
\put(2.90,0.5){\oval(0.4,0.5)[t]}
\put(3.30,0.5){\oval(0.4,0.5)[b]}
\put(3.70,0.5){\oval(0.4,0.80)[t]}
\put(4.40,0.5){\oval(1,1)[b]}
\put(5.10,0.5){\oval(0.4,0.5)[tl]}
\end{picture} \par
\vskip 0.75truecm
\caption{The symbolic representation of the dependence of the action
$S(U)$ on the gauge field configuration $\{ U_{x\mu} \}$.
The absolute minima correspond to double
Dirac sheets, and the local minima correspond to single
Dirac sheets.
There are also the local minima corresponding to two single
Dirac sheets, etc. (not shown).}
\label{fig:vacuum}
\end{figure}
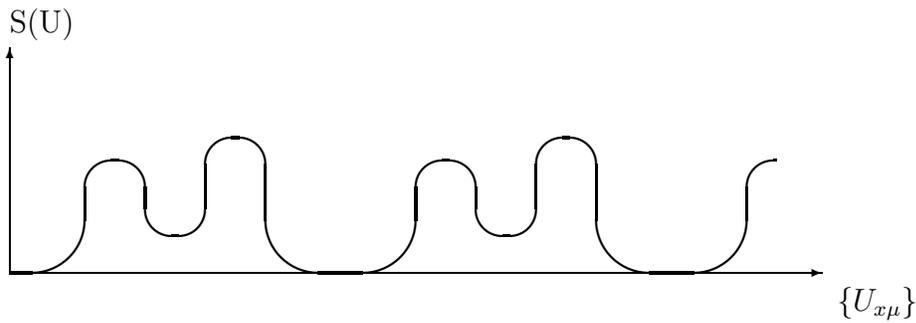

%
%
%
\begin{figure}[pt]
\vspace{-1.0cm}
\begin{center}
\leavevmode
\hbox{
\epsfysize=400pt\epsfbox{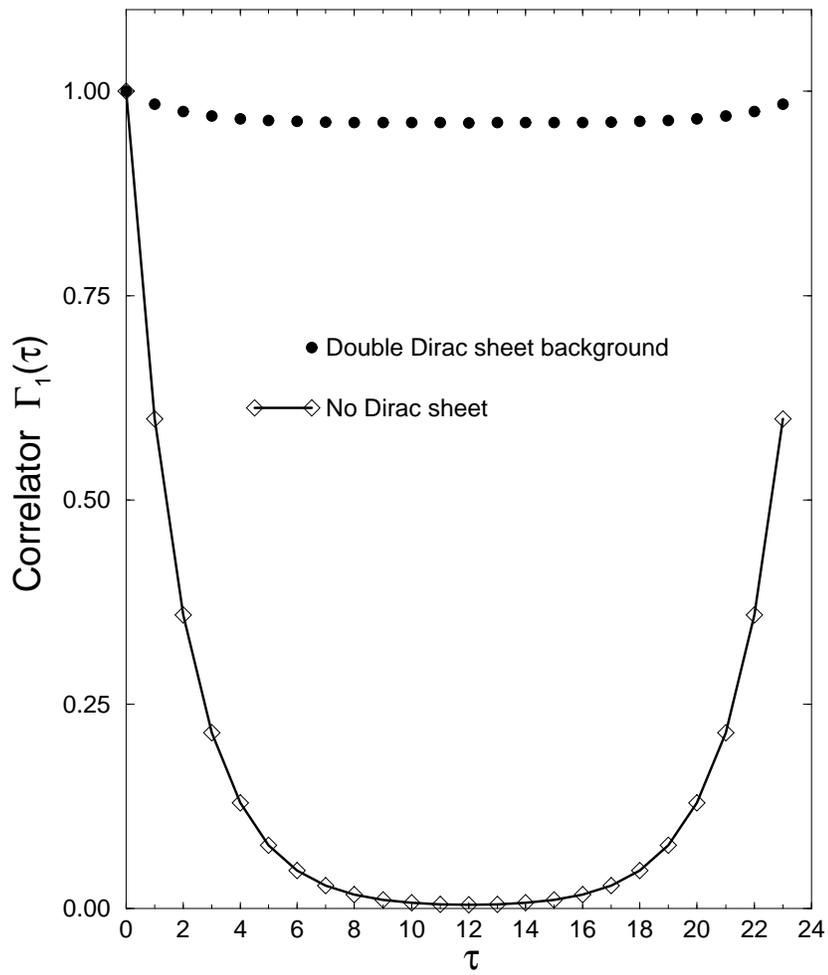}
     }
\end{center}
\vskip 0.25truecm
\caption{
The normalized correlators $\Gamma^{dds}_{1}(\tau ;{\vec p})$ 
(circles) and $\Gamma_1^{stan.}(\tau ;{\vec p})$ (diamonds)
on a $12^3\times 24$ lattice for $\vp =(0;\frac{\pi}{6};0)$ 
and $\beta =1.1$.
}
\label{fig:cor_3}
\vskip -0.2truecm
\end{figure}


\vspace{2cm}
\begin{thebibliography}{999}
\newcommand{\prd}[1]{Phys.~Rev.~{\bf D#1}\ }
\newcommand{\plb}[1]{Phys.~Lett.~{\bf #1B}\ }
\newcommand{\npb}[1]{Nucl.~Phys.~{\bf B#1}\ }
\newcommand{\prl}[1]{Phys.~Rev.~Lett.~{\bf #1}\ }
\newcommand{\prep}[1]{Phys.~Rep.~{\bf #1}\ }
\newcommand{\ap}[1]{Ann.~Phys.~{\bf #1}\ }
\newcommand{\cmp}[1]{Commun.~Math.~Phys.~{\bf #1}}
\newcommand{\rmp}[1]{Rev.~Mod.~Phys.~{\bf #1}}
\newcommand{\ptp}[1]{Prog.~Theor.~Phys.~{\bf #1}}
%
\bibitem{wil}    K. Wilson, \prd{10} (1974) 2445.
\bibitem{pol1}   A. M. Polyakov, Nucl. Phys. {\bf B120} (1977) 429.
\bibitem{grib}   V.N.~Gribov,           \npb{139} (1978) 1.
\bibitem{dgt}    T.A.~DeGrand and D.~Toussaint, \prd{22} (1980) 2478.
\bibitem{vkm}    V.K.~Mitrjushkin, HU Berlin--EP--96/21 and DESY 96--129.
\bibitem{neu}    V.~Gr\"osch, K.~Jansen, T.~Jers\'{a}k, C.B.~Lang,
                 T.~Neuhaus and C.~Rebbi, \plb{162} (1985) 171.
\bibitem{mon1}   J. S. Barber, R. E. Shrock and R. Schrader,
                 ~Phys. Lett. {\bf B152} (1985) 221. \\
                 J. S. Barber and R. E. Shrock,
                 \npb{257} [FS 14] (1985) 515.
\bibitem{mon2}   V.G. Bornyakov, V.K. Mitrjushkin and M. M\"uller--Preussker,\\
                 Nucl. Phys. {\bf B} (Proc. Suppl.) 30 (1993) 587.\\
                 A.~Hoferichter, V.K.~Mitrjushkin, M.~M\"uller\--Preussker,
                 Th.~Neuhaus and H.~St\"uben, Nucl.Phys. {\bf B434} (1995) 358.
\bibitem{nak}    A.~Nakamura and M.~Plewnia, \plb{255} (1991) 274.
\bibitem{bmmp}   V.G.~Bornyakov, V.K.~Mitrjushkin, M.~M\"uller-Preussker
                 and F. Pahl, \plb{317} (1993) 596.
\end{thebibliography}
\end{document}